# NETWORK SCIENCE
# A review focused on tourism

**Rodolfo Baggio**[1]
Bocconi University, Italy
**Noel Scott**
The University of Queensland, Australia
**Chris Cooper**
Oxford Brookes University, UK

**Abstract:** This paper presents a review of the methods of the science of networks with an application to the field of tourism studies. The basic definitions and computational techniques are described and a case study (Elba, Italy) used to illustrate the effect of network typology on information diffusion. A static structural characterization of the network formed by destination stakeholders is derived from stakeholder interviews and website link analysis. This is followed by a dynamic analysis of the information diffusion process within the destination demonstrating that stakeholder cohesion and adaptive capacity have a positive effect on information diffusion. The outcomes and the implications of this analysis for improving destination management are discussed.

INTRODUCTION

Understanding the shape and behavior of physical or social worlds requires an examination of the connections or relationships between elements of the phenomena under study and these connections may be represented as a network of links. The study of the structural and dynamic properties of such network representations of physical, biological, and social phenomena is called network science (Watts 2004). Network science utilizes a range of tools and techniques to examine how the topological or structural properties of a network affect its behavior or evolution. The topology of a network has been found to have a profound influence on its overall dynamic behavior and can be used to explain a wide number of processes, including the spread of viruses over a computer network and of diseases in a population; the formation of opinions and diffusion of information as well as the robustness of a system to external shocks. Network research has revealed that network behaviors and processes can be explained based upon the properties of a system's general connectivity and studies have found that the topology of many complex systems has been shown to share fundamental properties (Boccaletti, Latora, Moreno, Chavez and Hwang 2006).

In this paper tourism destinations are considered as complex systems, represented as a network by enumerating the stakeholders composing it and the linkages that connect them. While there is a significant literature on the importance of the relationships between tourists and service organizations and connecting tourism companies (Lazzeretti and Petrillo 2006; Morrison, Lynch and Johns 2004; Pavlovich 2003; Stokowski 1992; Tinsley and Lynch 2001), few works are available which examine a tourism destination from a network point of

---

view and fewer still that use quantitative methods of network science (Baggio 2008; Pforr 2006; Scott, Cooper and Baggio 2008b; Shih 2006).

The historical development of network science reveals a number of streams of thought (Scott, Cooper and Baggio 2007; Scott et al 2008a). The first is mathematically-based social network analysis which examines properties of "ideal" networks and is exemplified in the work of Burt (1992; 1997). A second stream uses qualitative methodology and is based in the social sciences, in which a network is viewed as an analogy for the interactions between individuals in a community. An example is the study of policy networks by Rhodes (1990; 1997). A third is the physicist's view of complex networks explored in the framework of statistical physics and complexity theory (Albert and Barabási 2002; Boccaletti et al 2006). While each of these three streams has advantages for the study of tourism, this paper focuses on the latter stream of thought. It aims to firstly, apply the quantitative methods of analysis of complex networks to the tourism field specifically focusing on understanding the tourism destination and thus secondly, to contribute to the methodological foundations of tourism (Tribe 1997).

**NETWORK SCIENCE**

A network is normally represented by a drawing in which the various elements are shown as dots and the connections among them as lines linking pairs of dots. This drawing, a mathematical abstraction, is called a graph and the branch of mathematics known as graph theory establishes the framework providing the formal language to describe it and its features. The application of networks in the social sciences using graphs and related social network analysis tools developed in the first half of $20^{th}$ century (Barnes 1952; Moreno 1934; Radcliffe-Brown 1940; Simmel 1908). The basic idea of this body of knowledge is that the structure of social interactions influences individual decisions, beliefs and behavior (Scott 2000). In this tradition, analyses are conducted on patterns of relationships rather than concentrating upon the attributes and behaviors of single individuals or organizations (Wasserman and Galaskiewicz 1994). By the end of the 1990s, the methods and possibilities of social network analysis were well established and formalized (Freeman 2004; Scott 2000; Wasserman and Faust 1994; Wellman and Berkowitz 1988), and network analysis had become a standard diagnostic and prescriptive tool in applied fields such as management and organization studies (Cross, Borgatti and Parker 2002; Haythornthwaite 1996; Tichy, Tushman and Fombrun 1979). These studies, while useful, tended to view a social system as static and were often criticized on the basis that they ignored the dynamic nature of organizations and groups.

Meanwhile scientists examining many natural and artificial systems had documented dynamic behavior that was non-linear and indeed exhibited complex or chaotic patterns over time. This led, in the second half of the $20^{th}$ century to detailed study and modeling of such nonlinear complex systems, facilitated by the power of modern computers albeit based upon ideas dating from the $18^{th}$ century (examples are: Euler 1736; Lyapunov 1892; Poincaré 1883; Strutt 1892). The consideration of the dynamic properties of networks began in the 1960s with the seminal work of Erdös and Rényi who presented a model of a random network (Erdös and Rényi 1959; 1960; 1961). The authors showed that dynamic growth in the number of connections gives rise to phenomena such as the formation of giant fully connected subnetworks, which seem to arise abruptly when some critical value of link density is attained. This finding attracted the interest of statistical physicists, well accustomed to analysis of these kinds of critical transitions in large systems. Three provocative papers (Barabási and Albert 1999; Faloutsos, Faloutsos and Faloutsos 1999; Watts and Strogatz 1998) in the late 1990s placed the analysis of networked systems in the context of statistical physics, providing a strong theoretical basis to these investigations, and justifying the search for universal properties of networked objects. The models proposed have made it possible to describe the static, structural and dynamic characteristics of a wide range of both natural and artificial complex networks and have highlighted the linkage between the topological properties and the functioning of a system, independent of the nature of the system's elements (Boccaletti et al 2006; Caldarelli 2007; Watts 2004). There is a growing literature applying these methods to the exploration of social and economic systems, driven by the interest in self-organizing processes and the emergence of ordered arrangements from randomness (Ball 2003; Castellano, Fortunato and Loreto 2009; Stauffer 2003).

*Complexity and Network Science: the theoretical framework*

There is no formal designation of a complex adaptive system despite a growing literature and debate by many. Instead, many authors characterize a system as complex and adaptive by listing the properties that these systems exhibit (see for example Cilliers 1998; Levin 2003; Ottino 2004). The most common and significant properties are:
- The system is composed of a large number of interacting elements;
- The interactions among the elements are nonlinear;
- Each element is unaware of the behavior of the system as a whole, it reacts only to locally available information;
- The system is usually open and in a state far from equilibrium; and
- Complex systems have a history, their actual and future behavior depend upon this history and are particularly sensitive to it.

Many real world ensembles are complex adaptive systems, as in economics where "even the simple models from introductory economics can exhibit dynamic behavior far more complex than anything found in classical physics or biology" (Saari 1995:222).

A tourism destination shares many of these characteristics, encompassing many different companies, associations, and organizations whose mutual relationships are typically dynamic and nonlinear (Michael 2003; Smith 1988). The response of stakeholders to inputs from the external world or from inside the destination may be largely unpredictable (Russell and Faulkner 2004). During the evolution of the destination system it is possible to recognize several reorganization phases in which new structures emerge such as the development of a coordinating regional tourism organization. Besides these "particular" or unique behaviors however, the system as a whole may also be found to follow general "laws". Models such as the one by Butler (1980), although discussed, criticized, amended and modified (Butler 2005a; b), are generally considered able to give meaningful descriptions of a tourism destination and, in many cases, have proved useful tools for managing destination development despite the peculiarities of individual case studies. More detailed studies can be found which have assessed the "complex" nature of tourism systems, both in a qualitative and a quantitative way (Baggio 2008; Farrell and Twining-Ward 2004; Faulkner and Russell 1997).

According to Amaral and Ottino (2004), the toolbox available to study such complex systems derives from three main areas of research: nonlinear dynamics, statistical physics and network science. First, research since the end of the 19$^{th}$ century has yielded several mathematical techniques which allow approximation of the solutions to the differential equations used to describe nonlinear systems that were non-solvable analytically. Today, the availability of powerful computers makes it possible to use numerical models and simulations to apply these techniques and thus chaotic and complex systems can be described in terms of the collective behaviors of their elementary components.

Second, research in statistical physics has provided macroscopic (statistical) approximations for the microscopic behaviors of large numbers of elements which constitute a complex system. In particular, it provides a theoretical foundation to the study of phase transitions (such as the one occurring to water in passing from liquid to solid or vapor) and the critical conditions governing them. Within a statistical physics framework, the analysis of data, the development and evaluation of models or the simulation of complex systems are carried out with the help of tools such as nonlinear time series analysis, cellular automata, and agent-based models (see Shalizi 2006 for an excellent review).

Two important concepts stem from this statistical physics tradition: universality and scaling (Amaral et al 2004). Universality is the idea that general properties, exhibited by many systems, are independent of the specific form of the interactions among their constituents, suggesting that findings in one type of system may directly translate into the understanding of many others. Scaling laws govern the variation of some distinctive parameters of a system with respect to its size. The mathematical expression of these laws applied to complex and chaotic systems involves a power law, now considered a characteristic signature of self-similarity.

The third area of research is based on the idea that a network can be used to represent many complex systems. The interactions among the different elements lead, in many cases, to global behaviors that are not observable at the level of the single elements, and they exhibit characteristics of emergence typical of a complex system. Moreover, their collective properties are strongly influenced by the topology of the linking network (Barabási 2002; Buchanan 2002). This is the approach followed in the rest of this paper.

*Characterization of Complex Networks*

The inter- and multi-disciplinary origin of network science has led to a wide variety of quantitative measurements of their topological characteristics (see da Fontoura Costa, Rodrigues, Travieso, and Villas Boas 2007 for a thorough review). Mathematically speaking, a network is represented by an ordered pair $G: = (V,E)$, where $V$ is a set whose elements are called vertices or nodes; $E$ is a set of pairs of distinct nodes, called edges or links. The graph can also be represented by a square *adjacency* matrix $A$. There is a full correspondence between a graph, a network and an adjacency matrix and the three terms are used indiscriminately. In particular, the identification between a graph and an adjacency matrix makes all the powerful methods of linear algebra available to a network scientist to investigate network characteristics. Table 1 provides the definition and the formulas for the main network metrics.

Table 1. Main network metrics

| Network metric | Description |
|---|---|
| *adjacency matrix* | square matrix whose elements $a_{x,y}$ have a value different from 0 if there is an edge from some node *x* to some node *y*. $a_{x,y} = 1$ *if the* link is a simple connection (unweighted graph). $a_{x,y} = w$ when the link is assigned some kind of weight (weighted graphs). If the graph is undirected (links connect nodes symmetrically), $A$ is a symmetric matrix. |
| *order* | total number of nodes: $n$ |
| *size* | total number of links: $m = \sum_i \sum_j a_{ij}$ |
| *nodal degree* | number of links connecting *i* to its neighbors: $k_i = \sum_i a_{ij}$ |
| *density* | the ratio between *m* and the maximum possible number of links that a |
| | graph may have: $\delta = \dfrac{2m}{n(n-1)}$; |
| *path* | a series of consecutive links connecting any two nodes in the network, the *distance* between two vertices is the length of the shortest path connecting them, the *diameter* of a graph is the longest distance (the maximum shortest path) existing between any two vertices in the graph: $D = \max(d_{ij})$, the *average path length* in the network is the arithmetical mean of all the distances: $l = \dfrac{1}{n(n-1)} \sum_{i \neq j} d_{ij}$. Numerical methods, such as the well known Dijkstra's algorithm (Dijkstra 1959) are used to calculate all the possible paths between any two nodes in a network. |
| *clustering coefficient* | the degree of concentration of the connections of the node's neighbors in a graph and gives a measure of local inhomogeneity of the link density. It is calculated as the ratio between the actual number $t_i$ of links connecting the neighborhood (the nodes immediately connected to a chosen node) of a node and the maximum possible number of links in that neighborhood: $C_i = \dfrac{2t_i}{k_i(k_i - 1)}$. For the whole network, the clustering coefficient is the arithmetic mean of the $C_i$: $C = \dfrac{1}{n} \sum_i C_i$; |
| *proximity ratio* | the ratio between clustering coefficient and average path length normalized to the values the same network would have in the hypothesis of a fully random distribution of links: $\mu = \dfrac{C/l}{C_{rand}/l_{rand}}$. It can be conceptualized as an index of small-worldness; |
| *efficiency* (at a global $E_{glob}$ or local $E_{loc}$ level) | measures the capability of the networked system (global) or of a single node (local) to exchange information. $E_{glob} = \dfrac{1}{n(n-1)} \sum_{i \neq j} \dfrac{1}{d_{ij}}$. $E_{loc,i} = \dfrac{1}{k_i(k_i-1)} \sum_{l \neq m} \dfrac{1}{d'_{lm}}$; for the whole network. Its average (called local efficiency of the network) is: $E_{loc} = \dfrac{1}{n} \sum_i E_{loc,i}$; |
| *assortative mixing coefficient* | gauges the correlation between the degrees of neighboring nodes. If positive, the networks are said to be assortative (otherwise disassortative). In an assortative network, well-connected elements (having high degrees) tend to be linked to each other. It is calculated as a Pearson correlation coefficient; $dg_i$ is the degree of node *i*, $dn_i$ the mean degree of its first neighbors: $r = \dfrac{\sum_i (dg_i - \overline{dg})(dn_i - \overline{dn})}{\sqrt{\sum_i (dg_i - \overline{dg})^2 \sum_i (dn_i - \overline{dn})^2}}$; the standard error can be calculated by using the bootstrap method (Efron and Tibshirani 1993). |

One important factor, found to be a strong characterizer of a network topology is the distribution of the degrees of its nodes. This is usually expressed as a statistical probability distribution *P(k)*, i.e. for each degree present in the network, the fraction of nodes having that degree is calculated. The empirical distribution is then plotted and a best fit functional (continuous) relationship describing it is determined. A cumulative version of the degree distribution *P(>k)* is also used. It gives the probability (fraction) of nodes having degree greater than a certain value (from the list of the values existing in the network).

A complex network exhibits, in many cases, some form of substructure. Local subgroups can have a "thickening" of within-group connections while having less dense linkages with nodes outside the group. The study of this modular structure of *communities* has attracted academic attention, since the existence of communities are a common characteristic of many real networked systems and may be central for the understanding of their organization and evolution. It may be possible, for example, to reveal social structure through communication patterns within a community. Different definitions of modularity exist and several methods have been proposed to measure it. They rely on numerical algorithms able to identify some topological similarity in the local patterns of linking (Arenas, Danon, Díaz-Guilera, Gleiser and Guimera 2004; Danon, Díaz-Guilera, Duch and Arenas 2005). In all of them, however, a quantity called *modularity index* is used to gauge the effectiveness of the outcomes (Clauset, Newman and Moore 2004; Girvan and Newman 2002). It is defined as: $Q = \sum_i (e_{ii} - a_i)^2$, where $e_{ii}$ is the fraction of edges in the network between any two vertices in the subgroup *i*, and $a_i$ the total fraction of edges with one vertex in the group. In other words, *Q* is the fraction of all edges that lie within a community minus the expected value of the same quantity in a graph in which the nodes have the same degrees but edges are placed at random. All of the metrics described in this section can be calculated with the help of standard software packages such as as Pajek (Batagelj and Mrvar 2007) or Ucinet (Borgatti, Everett and Freeman 1992).

*Network Models*

In a series of papers Erdös and Rényi (1959; 1960; 1961) propose a model (ER) in which a network is composed of a set of nodes and the links are placed randomly between pairs of nodes with probability *p*. The resulting degree distribution (in the limit of large numbers of nodes and links) follows a Poisson law with a peak *⟨k⟩* (the average degree of the network):

$$P(k) \approx \frac{\langle k \rangle^k}{k!} e^{-\langle k \rangle}.$$

The diameter, clustering coefficient and average path length of an ER network are proportional to the number of nodes and the probability *p*. The network also shows an interesting behavior when the connection probability increases. Over a certain critical threshold $p_c$, a very large group of connected nodes encompassing most if not all of the nodes (depending on the value of $p>p_c$), a *giant cluster*, forms. Below $p_c$ the network is composed of several disconnected subgraphs.

In the late 1990s, three influential papers (Barabási and Albert 1999; Faloutsos et al 1999; Watts and Strogatz 1998) presented empirical evidence of networks exhibiting topological characteristics different from those hypothesized by Erdös and Rényi. Watts and Strogatz (1998) discussed networks in which, contrary to what was expected from an ER model, the clustering coefficient was much higher, and, at the same time, the average path length remained small. They named these networks *small-world* (SW). In a small-world network, and as happens in many social networks, any two nodes are likely to be connected through a very short sequence of intermediate neighbors. Many examples of real world networks have this characteristic. Faloutsos et al (1999) and Barabási and Albert (1999) on the other hand, found evidence of networks having a degree distribution quite different from the random Poissonian ER distribution. Their networks exhibit a power-law scaling: *P(k) ~ k$^{-\gamma}$* with an exponent *γ > 1*. In other words, in their networks, a small fraction of nodes have a large number of immediate neighbors which are often called hubs, while a large number of nodes have a low degree. The Poissonian and Power law degree distributions for networks of the same order (1000 nodes) and size (3000 links) are shown in Figure 1.

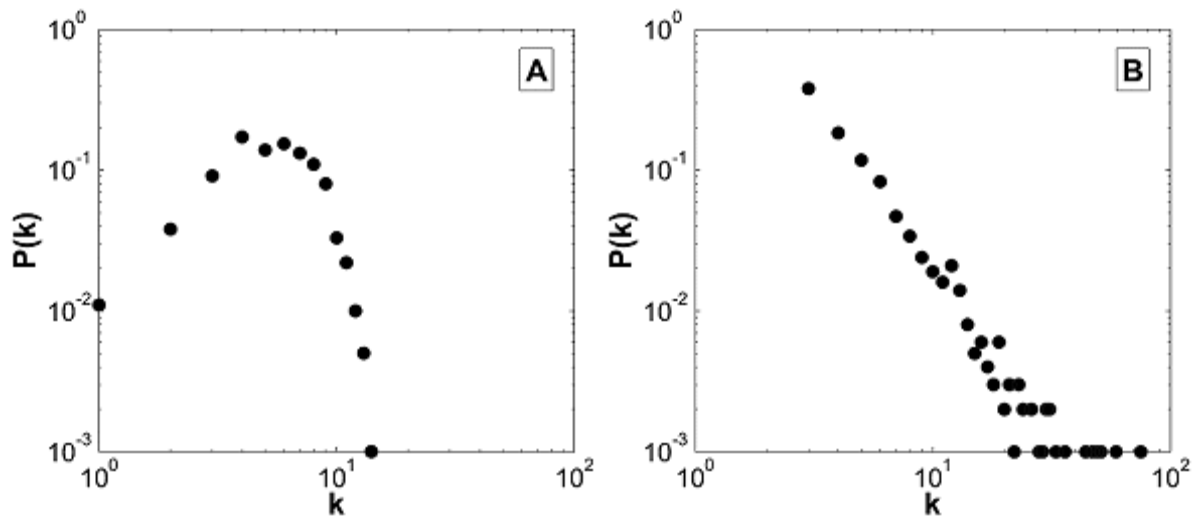

Figure 1. Degree Distributions: Poissonian (A) and Power-law (B)

These networks are called *scale-free* (SF) because they do not have a distinctive "scale"; (a typical number of connections per node) as is found in a Poissonian ER network in which the average (mean) degree characterizes the distribution. The SF model, first proposed by Barabási and Albert (1999) is a dynamic model. The power-law degree distribution is obtained if we consider a network as formed by adding nodes at successive time intervals, and adding links with a preferential attachment mechanism such that new nodes will connect with higher probability to nodes with high degrees (high number of links). This kind of rich-get-richer phenomenon has been observed in a large number of real networks, and there are several additions and modifications to account for the differences measured between the theoretical model and the real networks. Thus, we can modify the basic model by thinking of introducing a fitness parameter, which greatly increases the probability that a recent node has to be selected by the subsequent nodes; an aging limitation for which a node's capability to accept connections ends at a certain time interval (age); or an information constraint which puts a limit to the number of nodes among which a newcomer can select those to connect. Moreover, even in networks not growing by the addition of nodes, links can be added, deleted or moved (rewired) to adapt the network to specific conditions, and, thus besides the preferential attachment family, other mechanisms able to generate a power-law degree distribution exist (Albert et al. 2002; Bornholdt and Schuster 2002; Caldarelli 2007; Dorogovtsev and Mendes 2003; Durrett 2006; Li, Alderson, Tanaka, Doyle and Willinger 2005; Newman 2003b).

Mixed topologies have also been studied, both as abstract models (Mossa, Barthélémy, Stanley and Amaral 2002) and empirical observations (Baggio, Scott and Wang 2007;

Pennock, Flake, Lawrence, Glover and Giles 2002). The main characteristic of these networks is that they have a degree distribution which follows a power law for most part, but also has an inflecting or cut-off point. In statistical physics, power laws are associated with phase transitions (Landau and Lifshitz 1980; Langton 1990) or with fractal and self-similarity characteristics (Komulainen 2004). They also play a significant role for the description of those critical states between a chaotic and a completely ordered one, a condition known as self-organized criticality (Bak 1996; Bak, Tang and Wiesenfeld 1988). In other words finding a power law is one more confirmation of the "complexity" of the networked systems studied. As previously noted, many real networks exhibit scale-free properties. Tourism-related examples include the world-wide airport network (Guimerà and Amaral 2004), the websites of a tourism destination (Baggio 2007), the structural properties of interorganizational networks within destinations (Scott et al. 2008b), the paths followed by tourists reaching a destination by car (Shih 2006), or the world-wide flows of tourist arrivals (Miguéns and Mendes 2008). Many of these networks also exhibit small-world properties.

The wide variety of network models and empirical cases can be summarized following the classification proposed by Amaral, Scala, Bathélémy and Stanley (2000). These authors use the degree distribution *P(k)* to identify three broad classes of networks: single-scale exponential ER-like networks, scale-free networks and broad-scale networks with mixed types of degree distributions.

Besides the general depiction of the structural characteristics of the diverse networked systems presented, and beyond the different models and interpretations proposed, the literature on complex networks almost unanimously points out a strong relation between the topological structure and the functioning of the system described.

*Dynamic Processes*

A complex system is a dynamic entity. Economies, companies or tourism destinations can be thought of as living organisms existing in a state quite far from a static equilibrium. The only time in which they are in a full static equilibrium is when they are dead (Jantsch 1980; Ulgiati and Bianciardi 1997; Weekes 1995). In the literature, the growing interest in the development of models for a tourism destination (Butler 2005a; b), or the numerous methods devised to forecast some characteristic such as tourist demand (Song and Li 2008; Uysal and Crompton 1985; Witt and Witt 2000) are good testimonials of the dynamic nature of these systems and of the appeal of the study of these characteristics. As discussed above, the analysis of the topological properties of complex networks has provided interesting and useful outcomes as well as being intriguing from a theoretical point of view.

Growth processes have been studied for all the basic network types discussed in the previous section: the random (ER) graphs and the different types of scale-free networks. The behavior of a network with respect to possible disruptions (random or targeted removals of nodes and links) have been investigated and found to be strongly dependent on the network topology (Boccaletti et al 2006; Caldarelli 2007; Watts 2004).

One more important process is the diffusion process within a network and how it is influenced by the network topology. Epidemiological diffusion is a well-known phenomenon for which complete mathematical models have been devised (Hethcote 2000). It has long been known that the process shows a clearly defined threshold condition for the spread of an infection (Kermack and McKendrick 1927). This threshold depends on the density of the connections between the different elements of the network. However, this condition is valid only if the link distribution is random (as in an ER network). In some of the structured, non-homogeneous networks that make up the majority of real systems (e.g. SF networks), this threshold does not exist. Once initiated, the diffusion process unfolds over the whole network (Pastor-Satorras and Vespignani 2003).

*Methodological Issues and Epistemology*

There are two key issues to be considered in progressing network science and the study of tourism. The first of these is the epistemological legitimacy of applying the laws and methods of physics to a social activity such as tourism. The second relates to the practicalities of collecting data pertaining to a network. Applying the laws and methods of physics to a socio-economic system such as a tourism destination may raise an issue of epistemological legitimacy and is an area where there is little relevant prior literature. While a variety of works deal with these questions for both the natural and social sciences, and examine the attitudes and positions of researchers with regard to their approaches and methodologies (see for example Durlauf 1999; van Gigch 2002a; b), the specific problem of the applicability of a "physical" approach to social systems is little discussed and mostly only as a secondary topic. Physicists do not seem to feel the necessity to epistemologically justify their use of the knowledge and tools of physics in investigating other fields. Justifications and discussions are the job of the epistemologist and usually come very late in the development of a field of study. Certainly justifications are not considered necessary when, as in the case of network science, a discipline is still in a very early stage of development.

From a sociologist's perspective, however, the application of physical network theory may be rejected as irrelevant because it fails to address the recursive agency in the behavior of groups of people. Recursive agency refers to the ability of individuals to recognize their networked relationships and take proactive steps to change or modify their behavior. Thus, the applicability of "physical laws" governing human behavior is refused as non-applicable. One of the reasons for this refusal can be that a non-physicist has, sometimes, a mistaken idea of what physics is. Bernstein, Lebow, Stein and Weber (2000), for example, consider that sociologists mistakenly believe the ideas of physics are mainly those of Newtonian mechanics where single or small sets of particles are studied. Such particles have well defined characteristics (mass, velocity, energy) and, more importantly, their equations of motion can be described and investigated. Based on this idea, sociologists consequently object that a "social actor" is completely different from these homogeneous particles, as a social actor's behavior is influenced by their personal history, beliefs and personality and thus a system of particles is too simplistic a representation. If we consider models such as those proposed by Schelling (1971), Axelrod (1997) or Sznajd-Weron and Sznajd (2000) this remark seems justified.

However, physicists may have different aims from achieving such individual predictive outcomes. For example in studying a socio-economic system we may be focused on its global behavior and on the possibility of making predictions at a system level rather than seeking to predict the conduct of single elements (individual actors). This alternative aim seeks to understand how regularities may emerge (when they do) out of the apparently erratic behavior of single individuals (Majorana 1942). In this perspective, a comparison of theoretical predictions with empirical data has the primary objective of verifying whether the trends seen in the data are compatible with a "reasonable" conceptual modeling of the idealized actors and whether there is some level of consistency or additional factors are required to provide an explanation.

In these circumstances, as Castellano et al (2009) note, only high level characteristics, such as symmetries, critical transitions or conservation laws are relevant. These, as the findings of statistical physics show, do not depend on the individual details of the system but possess some universality characteristics. Thus if the aim is to examine such global properties, it is possible to "approach the modelization of social systems, trying to include only the simplest and most important properties of single individuals and looking for qualitative features exhibited by models" (Castellano et al 2009:2). These considerations lead us to justify the application of the laws and methods of statistical physics to the study of a socio-economic system such as a tourism destination, on the provision that the quantitative techniques rely on sound and accepted qualitative interpretations of the phenomena as described in this paper.

*Data Collection*

On many occasions full enumeration of data regarding a network (nodes and links) is not possible. This is especially true for social and economic systems, and is certainly the case for a tourism destination. It is possible to use sampling to study complex networks but this requires careful application. As long as we are considering a system in which the elements are placed at random, as in the case of an ER network, the "standard" statistical considerations can be made, and the significance of the sample assessed with standard methods (Cochran 1977). We have seen, however, in the previous section, that the effect of removing links or nodes from an inhomogeneous system such as an SF network can lead to dissimilar results and is "element dependent". We may easily imagine, then, that a sample of a network missing some critical hubs could lead us to wrong conclusions when examining its topology.

The literature on this subject is not extensive. The problem has been highlighted only as a consequence of the recent discoveries in the field. It has been found that in the case of a structured network (scale-free, for example) it is not possible to easily determine the significance of a sample collected. Depending on the results of the analysis of the data available, the researcher needs to make an educated guess of the final topology exhibited by the whole "population", i.e. the whole network. In the cases in which this is possible, then, we may determine how some of the main network metrics vary with the size of the sample and the topology of the network. In the case of an SF network (Kossinets 2006; Lee, Kim and Jeong 2006; Stumpf and Wiuf 2005), the degree distribution exponent and average path length decrease when nodes or links are sampled; the assortativity coefficient remains practically unchanged; the clustering coefficient decreases when nodes are sampled; and increases when links are sampled.

*A Case Study: a tourism destination*

The review above shows that a vast theoretical and empirical literature has been accumulated and shows network science to be an effective tool for understanding complex systems. The empirical study described in this section provides an example of the application of network analysis methods to a tourism destination - the island of Elba, Italy. Elba is a typical "sun and sand" destination in the Tyrrhenian Sea. Elba's economy depends mainly on the wealth generated by about half a million tourists spending some 3 million nights per year (data provided by Elba Tourist Board, 2008). After a long period of growth, Elba is experiencing a decline in the number of tourist arrivals. The organizations operating on the island are mainly small and medium family-run businesses. A lack of cooperation and an excessive 'independence' of the Elban tourism stakeholders is a problem highlighted by several studies (Pechlaner, Tallinucci, Abfalter and Rienzner 2003; Tallinucci and Testa 2006).

Elba was selected for study as it is geographically distinct, has accessible records concerning tourism actors and with a scale suitable for detailed examination. The core tourism organizations (hotels, travel agencies, associations, public bodies etc.), identified from the official local tourism board, form the nodes of the network. The connections among them were enumerated by consulting publicly available documents such as membership lists for associations and consortia, commercial publications, ownership and board of directors' records. The data obtained and its comprehensiveness were validated with a series of structured and unstructured interviews with a selected sample of local "knowledgeable informants" such as the directors of the local tourism board and of the main industrial associations, or consultants active in the area. These interviews revealed a very limited number of links that were not previously discovered and it seems reasonable to assume that the final layout is about 90% complete. All the links are considered undirected and of equal weight. The network thus obtained is depicted in Figure 2 along with its degree distribution [where $P(k)$ is the number of nodes having degree $k$].

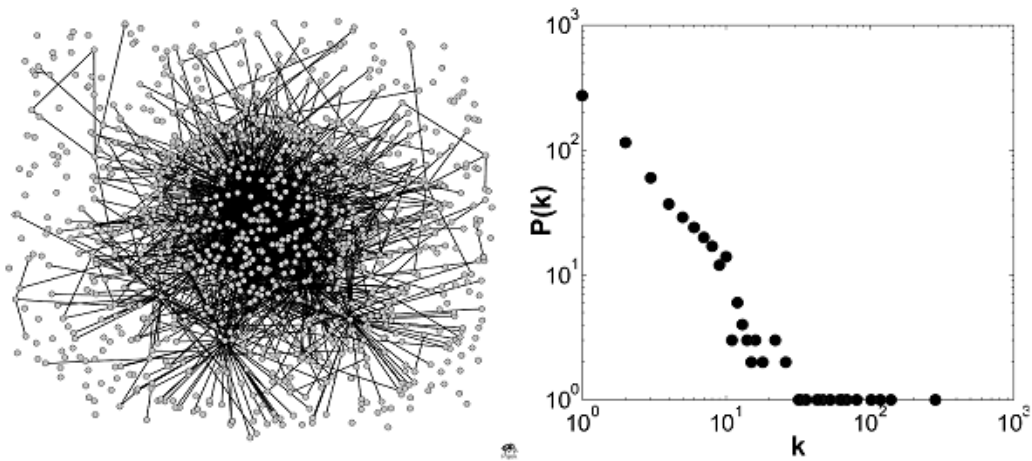

Figure 2. The Elba Destination Network and its Degree Distribution

Table 2. Elba Destination Network Metrics

| Metric | Elba network | Random network | Social networks | Web network |
|---|---|---|---|---|
| No. of nodes | 1028 | 1028 | | 468 |
| No. of links | 1642 | 1642 | | 495 |
| Density | 0.003 | 0.003 | $10^{-1} - 10^{-2}$ | 0.005 |
| Disconnected nodes | 37% | 3% | | 21% |
| Diameter | 8 | 13 | 10 | 10 |
| Average path length | 3.16 | 5.86 | 10 | 3.7 |
| Clustering coefficient | 0.05 | 0.003 | $10^{-1}$ | 0.014 |
| Degree distribution exponent | 2.32 | | | 2.17 |
| Proximity ratio | 34.09 | N/A | $10^2 - 10^3$ | 12.21 |
| Average degree | 3.19 | 3.25 | | 2.12 |
| Global efficiency | 0.131 | 0.169 | $10^{-1}$ | 0.17 |
| Local efficiency | 0.062 | 0.003 | $10^{-1}$ | 0.015 |
| Assortativity coefficient | -0.164 | 0.031 | $10^{-1}$ (>0) | -0.167 |

The results of the analysis of this network are summarized in Table 2. As a comparison, the second column contains the values calculated for a random (ER) network of the same order and size (the values are averages over 10 realizations). Table 2 also reports typical values for social networks published in the literature (see for example Albert et al. 2002; Boccaletti et al. 2006; Dorogovtsev and Mendes 2002; Newman 2003b).

The degree distribution for the Elban network (Figure 3) follows a power law $P(k) \sim k^{-\alpha}$. The exponent (and its standard error), calculated following the procedure proposed by Clauset, Shalizi and Newman (2009) is $\alpha = 2.32 \pm 0.27$.

The density of links is quite low, considering that the values found in the literature for the social networks studied are typically of the order of $10^{-1} – 10^{-2}$. Moreover, the percentage of nodes without connections is very high (39%). This results in a sparse network, also confirmed by the small value of the clustering coefficient. The efficiency of the Elban network is consequently quite low, both at a global and a local level. The assortativity coefficient is also different from what would have been expected in considering a socio-economic network such as Elba. This, as seen previously, represents the tendency of a node to connect with nodes having similar degrees. The correlation has been found positive for many of the social networks examined in the literature (Newman 2002), and, while debated by some authors (Whitney and Alderson 2006), this positivity is generally considered to be a distinguishing characteristic of social networks with respect to other systems. On the other hand, the calculated values for diameter and average path length seem to be in line with those of other real social systems and sensibly smaller than those exhibited by a random network. This indicates a certain level of compactness of the Elban network, at least for its central connected core. This is also confirmed by the proximity ratio which indicates a good level of "small-worldness" of the network.

The modularity of the network was calculated (Table 3) by dividing its actors with respect to the type of business (hospitality, associations, food and beverage services etc.) and geographical location (Elba's municipalities). As a comparison, the modularity was investigated using Clauset et al.'s (2004) algorithm which partitions the network on the basis of its connectivity characteristics, without supposing any division in advance (CNM in Table 3).

Table 3. Elba Network Modularity Analysis

| Grouping | No. of groups | Modularity | Average Modularity |
|---|---|---|---|
| Geography | 9 | 0.047 | 0.0052 |
| Type | 8 | -0.255 | -0.0319 |
| CNM | 11 | 0.396 | 0.0360 |
| CNM (random) | 12 | 0.367 | 0.0306 |

Table 3 shows the number of clusters identified (groups) and the modularity index. The last row reports (CNM random) the values calculated for a network of the same order and degree distribution as the Elban one with a randomized distribution of links (values are averages over 10 iterations). To better compare the different results, the last column of the table contains the average modularity over the groups (modularity/number of groups). All groups have a very low modularity. In one case (grouping by type), the negative value indicates that the actors tend to have more connections outside the group to which they belong than with businesses within the group. The higher values found by the CNM algorithm confirm that division by geography or by type of business does not imply any strong degree of clustering in these groups. The fact that the randomized network has a lower but similar modularity with respect to that obtained by using a community detection algorithm on the original network is an indication that a distinct modular structure exists even if not very well defined or highly significant (Guimerà, Sales-Pardo and Amaral 2004).

*The Topological Analogy: an example (real and virtual)*

Network science can also be applied to the virtual network among Elban tourism companies. The websites belonging to the tourism stakeholders were identified (only 'full' websites, with their own address were considered, discarding sets of pages embedded in the portals of other organizations) and the network (WN) was built by listing all the hyperlinks among them. This was accomplished by using a simple crawler and complementing the data obtained with a "manual" count of the hyperlinks to overcome the limitations of the program used (such as, for example, the impossibility of finding hyperlinks embedded in Flash applications or Java applets) (Baggio 2007). The last column in Table 2 shows the topological characteristics of the WN network compared with those of the "real" network described in the previous section.

As can be seen, apart from scale factors, most of the values have differences which are lower than an order of magnitude. Since in a complex network the distributions of these metrics are not normal, a simple comparison of their averages (arithmetic means) is an insufficient way of establishing similarities or dissimilarities. In these cases, as already proposed by some researchers (Clauset et al 2009; Leskovec and Faloutsos 2006), the Kolmogorov-Smirnov (KS) statistic is considered able to provide trustworthy results. The KS D-statistic gives the maximum distance between the cumulative probability distributions of empirical data $F(x)$ and $G(x)$ over the entire $x$ range: $D = \max_x |F(x) - G(x)|$. The statistic is nonparametric and insensitive to scaling issues, it compares only the shapes of the empirical distributions (Siegel and Castellan 1988).

The values for the D-statistics calculated when comparing the distributions of the web network with those of the real network are the following: degree = 0.119; clustering coefficient = 0.147; local efficiency = 0.125. As reference, the same values have been calculated for a random sample (RN) of the same size as WN, extracted from the real one. The values (averages over 10 realizations) are: degree = 0.147; clustering coefficient = 0.178; local efficiency = 0.184. The consistently lower values of the D-statistic in the case of the web network (with respect to the random sample) can be considered as a good confirmation of the likeness of their structural characteristics.

A strand of literature considers virtual networks as representations of the social relationships among the actors originating them. In essence: "computer networks are inherently social networks, linking people, organizations, and knowledge" (Wellman 2001:2031). Even if some argue that that the links are created in a rather unpredictable way, and it is not possible to find unambiguous meanings (Thelwall 2006), private or public organizations and companies consider a hyperlink as a strategic resource, and the structure of this network is created by specific communicative aims, rather than by accidental choices (Park and Thelwall 2003; Vaughan, Gao and Kipp 2006).

Based on these considerations and the network analysis, it is possible to formulate the following conjecture: the network of websites belonging to a cluster of (tourism) companies is a reliable sample of the whole socio-economic network formed by them. The obvious limitation is that the region examined must show a significant diffusion of the Internet and the Web. This, for a large part of the world, is not a severe limitation and thus the Web provides us with a relatively rapid, easy and objective way of sketching the main characteristics of such networks rather than more or less "randomly" sampling a socio-economic network with the usual investigation methods (Marsden 1990). The literature has produced much evidence on the issue of network sampling and the effect it might have on the topological characteristics of the whole network (Kossinets 2006; Lee et al. 2006). This must be taken into account in deriving the insights provided by network analysis methods.

*Dynamic Processes*

Networked systems, through their mathematical representation, are optimal candidates for numerical simulations. Indeed this technique is receiving increased attention as a powerful method to support complex analysis and planning activities for social and economic systems. Information and knowledge flows in a destination are important factors for the general "well-being" of the system. Efficiency, innovation and economic development are affected strongly by these processes. Moreover, the manner in which the diffusion unfolds influences the competitive advantage of individual actors and their planning of future actions (Argote and Ingram 2000).

A computer simulation can help assess the efficiency of information flows across the destination and test the capability of the system to react to some changes of its structural parameters. A simple epidemiological model can be employed. In this case, nodes are either "susceptible" to receiving information or already "infected" by it (i.e. they have received it). Despite its simplicity, this model is a reliable approximation (see for example Barthélemy, Barrat, Pastor-Satorras and Vespignani 2005; Xu, Wu and Chen 2007), and quite suitable to describe a knowledge transfer process. The simulation was conducted as follows: within a network, one randomly chosen stakeholder starts the spread by infecting a fraction $k_i$ of its immediate neighbors. At each subsequent time step, each infected element does the same until all the network nodes have been infected and the process ends. In this study, the model was run by adopting two different configurations. In the first case, the capacity of a stakeholder to transfer knowledge (spread infection) is used as a parameter for the model. It is defined as a probability $p(k_i)$ which determines the number of neighbors infected by a single actor. This justifies an important difference between the diffusion of information and knowledge and the spread of viruses. Viruses are indiscriminate, infecting any susceptible individual. Knowledge, on the other hand, is transferred only to a limited set of the individuals with which an actor has interactions (Huberman and Adamic 2004).

Particular actors, then, can have different "absorptive capacities" (Cohen and Levinthal 1990; Priestley and Samaddar 2007), i.e. different capabilities to acquire and retain the knowledge available to them due to the associated costs or their internal functioning, and to transfer it to other actors. In tourism, this issue is crucial for the high prevalence of small businesses that typically rely on external contacts for information. On the reasonable assumption that $p(k_i)$ depends on the size of the stakeholder, the network nodes were divided into three classes: large, medium and small (in our case we have the following proportions: large = 8%, medium = 17%, small = 75%). The values for $p(k_i)$ used in the simulations run are (arbitrarily) set as: $p(k_{large}) = 1$, $p(k_{medium}) = 0.8$, and $p(k_{small}) = 0.6$.

The second type of simulation aims at testing the influence of a network's structure, and particularly how the cohesion among stakeholders can affect the knowledge transfer process. In this case the experiment was performed with a modified version of the original network obtained by rewiring the connections while leaving unchanged the original connectivity (i.e. the number of immediate neighbors of each stakeholder and overall density of linkages), in order to obtain a higher clustering coefficient and a higher efficiency. The algorithm used is similar to the one proposed by Maslov and Sneppen (2002). The new network has a clustering coefficient $C = 0.274$ and a mean local efficiency $E_{loc} = 0.334$, as opposed to the original one whose values are $C = 0.084$ and $E_{loc} = 0.104$ (only the fully connected component of the Elban network was used, i.e. all isolated nodes were removed). As a comparison, a random network (same order and density, and random distribution of edges) was used. The time of peak diffusion, which can be used as an indicator of the process efficiency, decreases by 16% when comparing the random network with the Elban network containing different actors' capabilities. This, as expected, is due to the non-homogeneity of the network. When changing to equal capabilities (the original Elban network) a 22% improvement is found. A further consistent decrease (52%) is found when the local densities (clustering) are increased. These interventions have a significant impact on the information diffusion process. In other words: the spread of knowledge is faster if the network's connections are not distributed at random (scale-free in our case), it improves if all the stakeholders are considered to have equal absorptive capacities (the maximum) and is even more enhanced when the extent of formation of local groupings (collaborative communities) increases.

*Discussion*

The Elba tourism destination network has been characterized as a complex network whose main traits are common to many other natural and artificial systems. Its scale-freeness has been assessed. Despite this similarity, the structure differs from those exhibited by other complex systems mainly in its high degree of sparseness and very low degree of local clustering. In tourism terms this means that the local stakeholders exhibit a very low degree of collaboration or cooperation. A quantitative measurement for this feature is naturally derived from the metrics used for the network analysis. In particular, as argued elsewhere (Baggio 2007), the clustering coefficient (very low in this case) can be used as a measure of the extent of the degree of collaboration and the assortativity coefficient (very low and negative) can be thought of as representing the tendency to form collaborative groups. The qualitative knowledge of the destination (Pechlaner et al 2003; Tallinucci and Testa 2006) and the data gathered during the interviews conducted at the destination substantiate this interpretation. This apparent lack of collaboration among operators belonging to the same type has proved to be detrimental when thinking about the capacity of innovation which might help them to face the challenges of the contemporary highly competitive and globalized market. It has been shown, in fact, that a collaborative approach and intense exchanges, even in seemingly competitive organizations such as the group of Sydney hotels described by Ingram and Roberts (2000), may allow a valuable amalgamation of best practices, with the result of improving the performance and profitability of the whole group and its members. The low level of modularity unveiled further confirms this reading. It is interesting to note, in the results of the analysis that the highest modularity value is obtained with the usage of a "generic" numeric algorithm (Clauset et al 2004). This community structure, in the common understanding of the phenomenon (Arenas et al 2004), can be considered better than those which can be found based on the other criteria used: type of business and geographical location within the destination.

Moreover both the number and the composition of the clusters identified are different (Table 3). The system, in other words, exhibits self-organization properties which lead to the formation, to some extent, of an agglomeration of ties and produces a number of informal communities and an informal community structure. It can be concluded that the information contained in the geographical or business typology data does not represent fully the communality characteristics, and the modularity solutions found in this way are non optimal. This evidence has been also found in other social networks (Minerba, Chessa, Coppola, Mula and Cappellini 2007). From a destination management viewpoint, this result is important. It can provide indications on how to optimize destination performance by, for example, optimal communication pathways or even productivity in collaborations, overcoming rigid traditional subdivisions. It can provide a more practical tool to go along with the ideas and practices of an adaptive approach to the management of tourism destinations which has been advocated by some scholars (Farrell et al 2004).

A word of caution is necessary when considering extending the considerations made on network clustering and modularity to other cases. It has been shown, for example, that significant values for the clustering coefficient can also be accounted for by a simple random graph model (i.e. in which edges are placed at random), under the constraint of a fixed degree distribution $P(k)$. The emergence of this effect is a "statistical fluctuation" due to the form of the degree distribution in networks with a finite number of elements (Newman 2003a; Newman, Strogatz and Watts 2001). A correct interpretation of the result, therefore, can only be achieved by complementing the quantitative assessment with a deep knowledge of the social system under study, which typically comes from a tradition of qualitative investigations.

The worth of the methods presented here is well demonstrated by looking at the comparison made between the real and the virtual networks of the Elban tourism stakeholders. Even with the limitations discussed previously, it has been possible to formulate a conjecture – the similarity between the topologies of the two networks – which can prove extremely useful in speeding up and easing the process of collecting data to perform network analyses for socio-economic systems such as tourism destinations.

The information diffusion process analyzed provides us with some more important results. The simulated measurements of the diffusion speed confirm, first of all, the improvement in the efficiency of the whole process due to the existence of a structured network in place of a randomly linked system. Two conceptually different situations were simulated. The first one considered the stakeholders of the destination as elements with different capabilities to acquire and consequently retransmit information or knowledge. The second one assessed the effects of a change in the topology of the network obtained by optimizing it with respect to its efficiency. The results show a clear improvement in diffusion speed when all the actors are considered to have the same capacity to transfer information or knowledge. This is an important indication for a destination manager. Putting in place measures and actions aimed at reducing the differences in the absorptive capacities of the destination stakeholders can have a highly beneficial impact on the overall system. However, the results indicate that a similar effect, but with an even higher magnitude, can be obtained by optimizing network efficiency. The exchange of information among the nodes is much improved if the connectivity of the network is modified so as to increase the local efficiency, and consequently the clustering coefficient.

In other words, a very important determinant for the spread of knowledge in a socio-economic system such as a tourism destination is the presence of a structured topology in the network of relations that connect the different stakeholders, and more than that, the existence of a well-identified degree of local cohesion. This supports the notion that destination stakeholders should be encouraged to form clusters and to both compete and cooperate in order to exchange knowledge and hence to raise the overall competitiveness of the destination. Quantitative network methods can, therefore, not only assess this effect, but, more importantly, give practical indications on how to improve the process. By performing different simulations with different sets of initial parameters (distribution of absorptive capacities or different levels of clustering), it is possible to obtain different settings and evaluate the effects of the choice of parameters on the final result.

**CONCLUSION**

This paper has described the methods and the techniques that network science has assembled so far for the study of complex adaptive systems and as an example of their application, the case of a tourism destination has been discussed along with some implications of this approach. Taken alone, network analysis methods are undoubtedly an intriguing and intellectually stimulating exercise. Physicists know, however, that no matter how sophisticated and effective theoretical techniques can be, they have little value if applied to a phenomenon without coupling them with sound "physical interpretations". Translated into the language of social science, this means that a thorough knowledge of the object of analysis is crucial to obtain meaningful outcomes both from a theoretical and a practical point of view. This knowledge is the one provided by qualitative methods. As Gummesson points out: "by abolishing the unfortunate categories of qualitative/quantitative and natural sciences/social sciences that have been set against each other, and letting them join forces for a common goal – to learn about life – people open up for methodological creativity" (2007:226), therefore "qualitative and quantitative, natural and social are not in conflict but they should be treated in symbiosis" (2007:246).

In the twenty-first century, the strong focus on issues such as partnership, collaboration, cooperation and the benefits of the tools available for the investigation of the relationships between the elements of a socio-economic system have been discussed several times in the general management literature. The implications, it is argued, go well beyond the simple study of networks. These methods are recognized to have a strong potential to inform a wide number of concerns such as the use of technology, the study of epidemiological diffusion (from diseases to marketing or policy messages), the formation of consensual opinions and the impacts of these on organizational structure and performance (Parkhe, Wasserman and Ralston 2006).

In this respect, the methods of network science can prove highly beneficial in deepening the knowledge of the whole system and, coupled with more traditional procedures, can provide powerful tools to support those *adaptive* management practices considered by many the only practical way to steer the collective efforts of multiple organizations (Bankes 1993; Farrell et al 2004; Holling 1978; Ritter, Wilkinson and Johnston 2004).

The possibility of using quantitative techniques to analyze the relationships between tourism organizations opens new paths for the researcher interested in the structure, evolution, outcomes, effectiveness and the governance of the tourism system. This work, therefore, strongly supports the idea that triangulation of research methods can give the clues necessary to improve the analysis of tourism systems and their components (Davies 2003).

Further research in this area will first need to confirm the results obtained so far by increasing the number of examples studied. The methods employed in this paper clearly require some additional refinement both from a practical and a theoretical point of view. Moreover, the ever growing number of studies in network science, mainly from what concerns the dynamic evolution of a complex networked system, may suggest new models and new approaches which will need careful consideration for their applicability to the tourism field. As a final point, it is a firm conviction of the authors that a more rigorous establishment and adoption of methodological tools such as those used in this work, can be a powerful way to help tourism research transition towards a less *undisciplined* array of theories and models (Echtner and Jamal 1997; Tribe 1997).

## REFERENCES

Albert, R., and A. Barabási
    2002 Statistical Mechanics of Complex Networks. Review of Modern Physics 74:47-91.

Amaral, L., and J. Ottino
    2004 Complex Networks - Augmenting the Framework for the Study of Complex Systems. The European Physical Journal B 38:147-162.
Amaral, L., A. Scala, M. Barthélémy, and H. Stanley
    2000 Classes of Small World Networks. Proceedings of the National Academy of the Sciences of the USA 97:11149-11152.
Arenas, A., L. Danon, A. Díaz-Guilera, A. Gleiser, and R. Guimera
    2004 Community Analysis in Social Networks. The European Physical Journal B 38:373-380.
Argote, L., and P. Ingram
    2000 Knowledge Transfer: A Basis for Competitive Advantage in Firms. Organizational Behavior and Human Decision Processes 82:150–169.
Axelrod, R.
    1997 The Dissemination of Culture: A Model with Local Convergence and Global Polarization. The Journal of Conflict Resolution 41:203-226.
Baggio, R.
    2007 The Web Graph of a Tourism System. Physica A 379:727-734.
    2008 Symptoms of Complexity in a Tourism System. Tourism Analysis 13:1-20.
Baggio, R., N. Scott, and Z. Wang
2007 What Network Analysis of the WWW can tell us about the Organisation of Tourism Destinations. In Proceedings of the CAUTHE 2007 Conference. Sydney, Australia, 11-14 February.
Bak, P.
    1996 How Nature Works. The Science of Self-Organized Criticality. New York: Springer.
Bak, P., C. Tang, and K. Wiesenfeld
    1988 Self-Organized Criticality. Physical Review A 38:364-374.
Ball, P.
    2003 The Physical Modeling of Human Social Systems. Complexus 1:190-206.
Bankes, S.
    1993 Exploratory Modeling for Policy Analysis. Operations Research 41:435-449.
Barabási, A.
    2002 Linked: The New Science of Networks. Cambridge: Perseus.
Barabási, A., and R. Albert
    1999 Emergence of Scaling in Random Networks. Science 286:509-512.
Barnes, J.
    1952 Class and Committees in the Norwegian Island Parish. Human Relations 7:39-58.
Barthélemy, M., A. Barrat, R. Pastor-Satorras, and A. Vespignani
    2005 Dynamical patterns of epidemic outbreaks in complex heterogeneous networks. Journal of Theoretical Biology 235:275-288.
Batagelj, V., and A. Mrvar
    2007 Pajek - Program for Large Network Analysis. Sourced <http://pajek.imfm.si/>
Bernstein, S., R. Lebow, J. Stein, and S. Weber
    2000 God Gave Physics the Easy Problems: Adapting Social Science to an Unpredictable World. European Journal of International Relations 6:43-76.
Boccaletti, S., V. Latora, Y. Moreno, M. Chavez, and D.-U. Hwang
    2006 Complex Networks: Structure and Dynamics. Physics Reports 424:175-308.
Borgatti, S., M. Everett, and L. Freeman
    1992 UCINET. Harvard, MA: Analytic Technologies.
Bornholdt, S., and H. Schuster, eds.
    2002 Handbook of Graphs and Networks - From Genome to the Internet. Berlin: Wiley-VCH.
Buchanan, M.
    2002 Nexus: Small Worlds and the Ground-breaking Science of Networks. New York: Norton.
Burt, R.
    1992 Structural Holes: The Social Structure of Competition. Cambridge: Harvard University Press.
    1997 A Note on Social Capital and Network Content. Social Networks 19:355-373.
Butler, R.

    1980 The Concept of a Tourist Area Cycle of Evolution: Implications for Management of Resources. Canadian Geographer 24 5-12.
    2005a The Tourism Area Life Cycle, Vol. 1: Applications and Modifications. Clevedon: Channel View.
    2005b The Tourism Area Life Cycle, Vol. 2: Conceptual and Theoretical Issues. Clevedon: Channel View.
Caldarelli, G.
    2007 Scale-Free Networks: Complex Webs in Nature and Technology. Oxford: Oxford University Press.
Castellano, C., S. Fortunato, and V. Loreto
    2009 Statistical Physics of Social Dynamics. Reviews of Modern Physics 81:591-646.
Cilliers, P.
    1998 Complexity and Postmodernism: Understanding Complex Systems. London: Routledge.
Clauset, A., M. Newman, and C. Moore
    2004 Finding Community Structure in Very Large Networks. Physical Review E 70:066111.
Clauset, A., C. Shalizi, and M. Newman
    2009 Power-law Distributions in Empirical Data. SIAM Review 51: 661-703.
Cochran, W.
    1977 Sampling Techniques. New York: John Wiley.
Cohen, W., and D. Levinthal
    1990 Absorptive Capacity: A New Perspective on Learning and Innovation. Administrative Science Quarterly 35:128-152.
Cross, R., S. Borgatti, and A. Parker
    2002 Making Invisible Work Visible: Using Social Network Analysis to Support Strategic Collaboration. California Management Review 44(2):25-46.
da Fontoura Costa, L., A. Rodrigues, G. Travieso, and P. Villas Boas
    2007 Characterization of Complex Networks: A Survey of Measurements. Advances in Physics 56:167-242.
Danon, L., A. Díaz-Guilera, J. Duch, and A. Arenas
    2005 Comparing Community Structure Identification. Journal of Statistical Mechanics: P09008
Davies, B.
    2003 The Role of Quantitative and Qualitative Research in Industrial Studies of Tourism. International Journal of Tourism Research 5:97-111.
Dorogovtsev, S., and J. Mendes
    2002 Evolution of Networks. Advances in Physics 51:1079-1187.
    2003 Evolution of Networks: From Biological Nets to the Internet and WWW. Oxford: Oxford University Press.
Durlauf, S.
    1999 How can Statistical Mechanics contribute to Social Science? Proceedings of the National Academy of the Sciences of the USA 96:10582-10584.
Durrett, R.
    2006 Random Graph Dynamics. Cambridge: Cambridge University Press.
Echtner, C., and T. Jamal
    1997 The Disciplinary Dilemma of Tourism Studies. Annals of Tourism Research 24:868-883.
Erdös, P., and A. Rényi
    1959 On Random Graphs. Publicationes Mathematicae (Debrecen) 6:290-297.
    1960 On the Evolution of Random Graphs. Publications of the Mathematical Institute of the Hungarian Academy of Sciences 5:17-61.
    1961 On the Strength of Connectedness of a Random Graph. Acta Mathematica Academiae Scientiarum Hungaricae 12:261-267.
Euler, L.
    1736 Solutio Problematis ad Geometriam Situs Pertinentis. Commentarii Academiae Scientiarum Imperialis Petropolitanae 8:128-140.
Faloutsos, M., P. Faloutsos, and C. Faloutsos

    1999 On Power-Law Relationships of the Internet Topology. Computer Communication Review 29:251-262.
Farrell, B., and L. Twining-Ward
    2004 Reconceptualizing Tourism. Annals of Tourism Research 31:274-295.
Faulkner, B., and R. Russell
    1997 Chaos and Complexity in Tourism: In Search of a New Perspective. Pacific Tourism Review 1:93-102.
Freeman, L
    2004 The Development of Social Network Analysis: A Study in the Sociology of Science. Vancouver: Empirical Press.
Girvan, M., and M. Newman
    2002 Community Structure in Social and Biological Networks. Proceedings of the National Academy of the Sciences of the USA 99:7821-7826.
Guimerà, R., and L. Amaral
    2004 Modeling the World-Wide Airport Network. The European Physical Journal B 38:381-385.
Guimerà, R., Sales-Pardo, M. and Amaral, L.
    2004 Modularity from fluctuations in random graphs and complex networks, Physical Review E 70:025101(R)
Gummesson, E.
    2007 Case Study Research and Network Theory: Birds of a Feather. Qualitative Research in Organizations and Management 2:226-248.
Haythornthwaite, C.
    1996 Social Network Analysis: An Approach and Technique for the Study of Information Exchange. Library & Information Science Research 18:323-342.
Hethcote, H.
    2000 The Mathematics of Infectious Diseases SIAM Review 42:599-653.
Holling, C., ed.
    1978 Adaptive Environmental Assessment and Management. New York: John Wiley and Sons.
Huberman, B., and L. Adamic
    2004 Information Dynamics in a Networked World. In Complex networks. Lecture Notes in Physics, Vol. 650, E. Ben-Naim, H. Frauenfelder and Z. Toroczkai, eds., pp. 371-398. Berlin: Springer-Verlag.
Ingram, P., and P. Roberts
    2000 Friendships among Competitors in the Sydney Hotel Industry. American Journal of Sociology 106:387–423.
Jantsch, E.
    1980 The Self-Organizing Universe. New York: Pergamon Press.
Kermack, W., and A. McKendrick
    1927 Contributions to the Mathematical Theory of Epidemics, Part 1. Proceedings of the Royal Society of London A 115 700-721.
Komulainen, T.
    2004 Self-Similarity and Power Laws. In Complex Systems - Science on the Edge of Chaos (Report 145, October 2004), H. Hyötyniemi, ed. Helsinki: Helsinki University of Technology, Control Engineering Laboratory.
Kossinets, G.
    2006 Effects of Missing Data in Social Networks. Social Networks 28:247-268.
Landau, L, and E. Lifshitz
    1980 Statistical Physics - Part 1. Oxford: Pergamon Press.
Langton, C.
    1990 Computation at the Edge of Chaos: Phase Transitions and Emergent Computation. Physica D 42:12-37.
Lazzeretti, L., and C. Petrillo, eds.
    2006 Tourism Local Systems and Networking. Amsterdam: Elsevier.
Lee, S., P. Kim, and H. Jeong
    2006 Statistical Properties of Sampled Networks. Physical Review E 73:016102

Leskovec, J., and C. Faloutsos
　2006 Sampling from Large Graphs. In 12th ACM SIGKDD International Conference on Knowledge Discovery and Data Mining pp. 631-636. Philadelphia, PA, USA (August 20-23).
Levin, S.
　2003 Complex Adaptive Systems: Exploring the Known, the Unknown and the Unknowable. Bulletin of the American Mathematical Society 40:3-19.
Li, L., D. Alderson, R. Tanaka, J. Doyle, and W. Willinger
　2005 Towards a Theory of Scale-free Graphs: Definition, Properties, and Implications. Internet Mathematics 2:431-523.
Lyapunov, A.
　1892 General Problem on Motion Stability. Kharkov: Kharkovskoye Matematicheskoe Obshchestvo (vol. 11, in Russian).
Majorana, E.
　1942 Il Valore delle Leggi Statistiche nella Fisica e nelle Scienze Sociali. Scientia 71:58-66.
Marsden, P.
　1990 Network Data and Measurement. Annual Review of Sociology 16:435-463.
Maslov, S., and K. Sneppen
　2002 Specificity and Stability in Topology of Protein Networks. Science 296:910-913.
Michael, E.
　2003 Tourism Micro-clusters. Tourism Economics 9:133-145.
Miguéns, J., and J. Mendes
　2008 Travel and Tourism: Into a Complex Network. Physica A 387:2963-2971.
Minerba, L., A. Chessa, R.C. Coppola, G. Mula, and G. Cappellini
　2007 A Complex Network Analysis of a Health Organization. Igiene e Sanità Pubblica
Moreno, J.
　1934 Who Shall Survive? Washington, DC: Nervous and Mental Disorders Publishing Co.
Morrison, A., P. Lynch, and N. Johns
　2004 International Tourism Networks. International Journal of Contemporary Hospitality Management 16:197-202.
Mossa, S., M. Barthélémy, H. Stanley, and L. Amaral
　2002 Truncation of Power Law Behavior in "Scale-Free" Network Models due to Information Filtering. Physical Review Letters 88:138701.
Newman, M.
　2002 Assortative Mixing in Networks. Physical Review Letters 89:208701.
　2003a Random Graphs as Models of Networks. In Handbook of Graphs and Networks, S. Bornholdt and H.G. Schuster, eds., pp. 35–68. Berlin: Wiley-VCH.
　2003b The Structure and Function of Complex Networks. SIAM Review 45:167-256.
Newman, M., S. Strogatz, and D. Watts
　2001 Random Graphs with Arbitrary Degree Distributions and their Applications. Physical Review E 64:026118.
Ottino, J.
　2004 Engineering Complex Systems. Nature 427:399.
Park, H., and M. Thelwall
　2003 Hyperlink Analyses of the World Wide Web: A Review. In Journal of Computer Mediated Communication [On-line]. Sourced <http://jcmc.indiana.edu/vol8/issue4/park.html>
Parkhe, A., S. Wasserman, and D. Ralston
　2006 New Frontiers in Network Theory Development. The Academy of Management Review 31:560-568.
Pastor-Satorras, R., and A. Vespignani
　2003 Epidemics and Immunization in Scale-free Networks. In Handbook of Graphs and Networks, S. Bornholdt and H. Schuster, eds. Berlin: Wiley-VCH.
Pavlovich, K.
　2003 Pyramids, Pubs, and Pizzas: An Interpretation of Tourism Network Structures. Tourism Culture & Communication 4:41-48.
Pechlaner, H., V. Tallinucci, D. Abfalter, and H. Rienzner

  2003 Networking for Small Island Destinations – The Case of Elba. In Information and Communication Technologies in Tourism, A. Frew, M. Hitz and P. O'Connor, eds., pp. 105-114. Wien: Springer.
Pennock, D., G. Flake, S. Lawrence, E. Glover, and C. Giles
  2002 Winners Don't Take All: Characterizing the Competition for Links on the Web. Proceedings of the National Academy of the Sciences of the USA 99: 5207-5211.
Pforr, C.
  2006 Tourism Policy in the Making: An Australian Network Study. Annals of Tourism Research 33:87-108.
Poincaré, H.
  1883 Sur certaines solutions particulières du problème des trois corps. Comptes Rendus de l'Académie des Sciences, Paris 97: 251-252.
Priestley, J., and S. Samaddar
  2007 Multi-Organizational Networks: Three Antecedents of Knowledge Transfer. International Journal of Knowledge Management 3:86-99.
Radcliffe-Brown, A.
  1940 On Social Structure. The Journal of the Royal Anthropological Institute of Great Britain and Ireland 70:1-12.
Rhodes, R.
  1990 Policy Networks: A British Perspective. Journal of Theoretical Politics 2:293-317.
  1997 Understanding Governance. Policy Networks, Governance, Reflexivity and Accountability. Buckingham: Open University Press.
Ritter, T., I. Wilkinson, and W. Johnston
  2004 Managing in Complex Business Networks. Industrial Marketing Management 33 175-183.
Russell, R., and B. Faulkner
  2004 Entrepreneurship, Chaos and the Tourism Area Lifecycle. Annals of Tourism Research 31:556-579.
Saari, D.
  1995 Mathematical Complexity of Simple Economics. Notices of the American Mathematical Society 42:222-230.
Schelling, T.
  1971 Dynamic Models of Segregation. Journal of Mathematical Sociology 1:143-186.
Scott, J.
  2000 Social Network Analysis: A Handbook. London: Sage Publications.
Scott, N., R. Baggio, and C. Cooper
  2008a Network Analysis and Tourism: From Theory to Practice. Clevedon, UK: Channel View.
Scott, N., C. Cooper, and R. Baggio
  2007 Use of network analysis in tourism research. In Advances in Tourism Marketing Conference (ATMC). Valencia, Spain, 10-12 September.
  2008b Destination Networks - Theory and Practice in Four Australian Cases. Annals of Tourism Research 35:169-188.
Shalizi, C.
  2006 Methods and Techniques of Complex Systems Science: An Overview. In Complex Systems Science in Biomedicine, T.S. Deisboeck and J.Y. Kresh, eds., pp. 33-114. New York: Springer.
Shih, H.
  2006 Network Characteristics of Drive Tourism Destinations: An Application of Network Analysis in Tourism. Tourism Management 27:1029-1039.
Siegel, S., and N. Castellan
  1988 Nonparametric Statistics for the Behavioral Sciences. New York: McGraw-Hill.
Simmel, G.
  1908 Soziologie. Berlin: Dunker and Humblot.
Smith, S.
  1988 Defining Tourism, A Supply-Side View. Annals of Tourism Research 15:179-190.
Song, H., and G. Li

2008 Tourism Demand Modelling and Forecasting - A Review of Recent Research. Tourism Management 29 203-220.
Stauffer, D.
  2003 Sociophysics Simulations. Computing in Science and Engineering 5:71-75.
Stokowski, P.
  1992 Social Networks and Tourist Behavior. American Behavioral Scientist 36:212-221.
Strutt, J., Lord Rayleigh
  1892 On the Instability of a Cylinder of Viscous Liquid under Capillary Force. Philosophical Magazine 34:145-154.
Stumpf, M., and C. Wiuf
  2005 Sampling Properties of Random Graphs: The Degree Distribution. Physical Review E 72:036118.
Sznajd-Weron, K., and J. Sznajd
  2000 Opinion Evolution in Closed Community. International Journal of Modern Physics C 11:1157-1165.
Tallinucci, V., and M. Testa
  2006 Marketing per le Isole. Milano Franco Angeli.
Thelwall, M.
  2006 Interpreting Social Science Link Analysis Research: A Theoretical Framework. Journal of the American Society for Information Science and Technology 57:60-68.
Tichy, N., M. Tushman, and C. Fombrun
  1979 Social Network Analysis for Organizations. The Academy of Management Review 4:507-519.
Tinsley, R., and P. Lynch
  2001 Small Tourism Business Networks and Destination Development. Hospitality Management 20:367-378.
Tribe, J.
  1997 The Indiscipline of Tourism. Annals of Tourism Research 24:638-657.
Ulgiati, S., and C. Bianciardi
  1997 Describing States and Dynamics in Far from Equilibrium Systems. Needed a Metric within a System State Space. Ecological Modelling 96:75-89.
Uysal, M., and J. Crompton
  1985 An Overview of Approaches to Forecast Tourism Demand. Journal of Travel Research 23:7-15.
van Gigch, J.
  2002a Comparing the Epistemologies of Scientific Disciplines in Two Distinct Domains: Modern Physics Versus Social Sciences. I: The Epistemology and Knowledge Characteristics of the Physical Sciences. Systems Research and Behavioral Science 19:199-209.
  2002b Comparing the Epistemologies of Scientific Disciplines in Two Distinct Domains: Modern Physics Versus Social Sciences. II: Epistemology and Knowledge Characteristics of the 'New' Social Sciences. Systems Research and Behavioral Science 19:551-562.
Vaughan, L., Y. Gao, and M. Kipp
  2006 Why are Hyperlinks to Business Websites Created? A Content Analysis. Scientometrics 67:291-300.
Wasserman, S., and K. Faust
  1994 Social Network Analysis. Methods and Applications. Cambridge, MA: Cambridge University Press.
Wasserman, S., and J. Galaskiewicz
  1994 Advances in Social Network Analysis: Research in the Social and Behavioral Sciences. Thousand Oaks, CA: Sage.
Watts, D.
  2004 The 'New' Science of Networks. Annual Review of Sociology 30:243-270.
Watts, D., and S. Strogatz
  1998 Collective Dynamics of 'Small World' Networks. Nature 393:440-442.

Weekes, W.
    1995 Living Systems Analysis and Neoclassical Economics: Towards a New Economics Paradigm. Systems Practice 8:107-115.
Wellman, B.
    2001 Computer Networks as Social Networks. Science 293:2031-2034.
Wellman, B., and S.D. Berkowitz, eds.
    1988 Social Structures: A Network Approach. Cambridge: Cambridge University Press.
Whitney, D., and D. Alderson
    2006 Are Technological and Social Networks Really Different? *In* Sixth International Conference on Complex Systems (ICCS2006). Boston, MA, June, 25-30.
Witt, S., and C. Witt
    2000 Forecasting Tourism Demand: A Review of Empirical Research. *In* The Economics of Tourism, C.A. Tisdell, ed., pp. 141-169. Cheltenham, UK: Edward Elgar Publishing Ltd.
Xu, X., Z. Wu, and G. Chen
    2007 Epidemic Spreading in Lattice-embedded Scale-free Networks. Physica A 377:125-130.